\documentclass[sigconf]{acmart}
\usepackage{sigir26-uww}
\begin{document}
\title{Understanding Wacky Weights: A Dissection of SPLADE's Learned Term Importance}

\author{Gregory Polyakov}
\email{grigorii.poliakov@uni-tuebingen.de}
\affiliation{%
  \institution{University of T{\"u}bingen}
  \city{Tübingen}
  \country{Germany}
}

\author{Harrisen Scells}
\email{harrisen.scells@uni-tuebingen.de}
\affiliation{%
  \institution{University of T{\"u}bingen}
  \city{Tübingen}
  \country{Germany}
}

\author{Carsten Eickhoff}
\email{carsten.eickhoff@uni-tuebingen.de}
\affiliation{%
  \institution{University of T{\"u}bingen}
  \city{Tübingen}
  \country{Germany}
}

\begin{abstract}
Learned sparse retrieval models such as SPLADE combine the effectiveness of neural architectures with the efficiency of inverted indices. As these models assign weights to terms from a fixed vocabulary, interpretability is often touted as a major benefit of these models. However, the emergence of wacky weights, i.e., expansion terms that appear semantically unrelated to the input, limits interpretability. While prior research has anecdotally observed this phenomenon, there is a lack of systematic understanding regarding their origins, prevalence, and contribution to retrieval effectiveness. In this paper, we reproduce SPLADE-v2 to systematically investigate wacky weights across the SPLADE family of models. We present a comprehensive dissection of wacky weights, providing a formal definition of wackiness based on the lexical utility of expansion terms. Furthermore, we introduce a novel measure to compare the prevalence of these tokens across models with varying vocabularies and sparsity levels. Beyond reproducing the original SPLADE-v2, we train it with various loss functions, datasets, and backbone transformers to isolate the factors contributing to wackiness. Our results show that larger vocabularies are associated with a higher prevalence of wacky tokens, while stricter sparsity regularizers are associated with lower prevalence. Finally, we find that wacky weights are used primarily for in-domain effectiveness rather than out-of-domain generalization.%
\footnote{Code: \url{https://github.com/polgrisha/understanding-wacky-weights}}
\end{abstract}

\ccsdesc[500]{Information systems~Retrieval models and ranking}
\ccsdesc[300]{Information systems~Language models}

\keywords{learned sparse retrieval, interpretability, wacky weights}

\copyrightyear{2026}
\acmYear{2026}
\setcopyright{cc}
\setcctype{by}
\acmConference[SIGIR '26]{Proceedings of the 49th International ACM SIGIR Conference on Research and Development in Information Retrieval}{July 20--24, 2026}{Melbourne, VIC, Australia}
\acmBooktitle{Proceedings of the 49th International ACM SIGIR Conference on Research and Development in Information Retrieval (SIGIR '26), July 20--24, 2026, Melbourne, VIC, Australia}
\acmDOI{10.1145/3805712.3808562}
\acmISBN{979-8-4007-2599-9/2026/07}

\maketitle

\section{Introduction}
Learned sparse retrieval (LSR) models are a compelling alternative to dense retrievers and cross-encoders, providing a trade-off between the effectiveness of transformer architecture and the efficiency of inverted-index search. LSR models such as SPLADE~\cite{formal2021splade, lassance2024splade} enable this trade-off by representing queries and documents as high-dimensional sparse vectors. Through the training process, LSR models learn to suppress spurious terms such as stop words and expand inputs with useful ones such as synonyms. A key advantage of these models is the interpretability afforded by the expansion terms, which contrasts the opaque vector outputs of dense retrieval and the black-box scoring of cross-encoders by providing human-readable tokens with explicit weights. 

\begin{figure}
\begin{tikzpicture}[
	node distance=.05cm and 0cm,
	capsule/.style={rounded corners,fill=green-300,text width=\columnwidth,minimum height=.45cm,align=center,inner sep=0cm},
	model/.style={text width=2.2cm,minimum height=1.35cm},
]
\node[draw,rounded corners,fill=gray-100, text width=.75\columnwidth] (query) {\enskip\faSearch\enskip{}symptoms of liver distress};

\node[model, below=.85cm of query,label={[align=center,text width=2cm,minimum height=.75cm]above:\textbf{SPLADE-L1}}] (model2) {\tikz{
\node[capsule,color=gray-100,fill=red-700] (t1) {feeling};
\node[capsule,below=of t1,fill=yellow-600] (t2) {bird};
\node[capsule,below=of t2,fill=yellow-600] (t3) {cause};
\node[below=0cm of t3,inner sep=0cm,minimum height=.25cm] (t4) {...};
\node[capsule,below=0cm of t4,fill=yellow-500] (t5) {hospital};
\node[capsule,below=of t5,fill=yellow-500] (t6) {distressed};
\node[capsule,below=of t6,fill=yellow-400] (t7) {panic};
}};

\node[model, left=of model2,label={[align=center,text width=2cm,minimum height=.75cm]above:\textbf{BM25+RM3}}] (model1) {\tikz{
\node[capsule, color=gray-100, fill=red-800] (t1) {respiratori};
\node[capsule,below=of t1, color=gray-100,  fill=red-800] (t2) {includ};
\node[capsule,below=of t2,fill=yellow-600] (t3) {mai};
\node[below=0cm of t3,inner sep=0cm,minimum height=.25cm] (t4) {...};
\node[capsule,below=0cm of t4,fill=yellow-400] (t5) {caus};
\node[capsule,below=of t5,fill=green-300] (t6) {gastrointestin};
\node[capsule,below=of t6,fill=green-100] (t7) {jaundic};
}};

\node[model,right=of model2,label={[align=center,text width=2cm,minimum height=.75cm]above:\textbf{SPLADE-\\DistilRoBERTa}}] (model3) {\tikz{
\node[capsule,color=gray-100,fill=red-900] (t1) {."};
\node[capsule,below=of t1,color=gray-100,fill=red-900] (t2) {\textvisiblespace{}L};
\node[capsule,below=of t2,color=gray-100,fill=red-900] (t3) {\textvisiblespace{}She};
\node[below=0cm of t3,inner sep=0cm,minimum height=.25cm] (t4) {...};
\node[capsule,below=0cm of t4,fill=yellow-500] (t5) {\textvisiblespace{}cause};
\node[capsule,below=of t5,fill=yellow-400] (t6) {\textvisiblespace{}depression};
\node[capsule,below=of t6,fill=yellow-400] (t7) {\textvisiblespace{}signs};
}};

\node[right=.3cm of model3,minimum width=.05\columnwidth,inner sep=0cm,rounded corners=.2cm,clip,label={[rotate=270,anchor=center,yshift=.3cm,color=gray-1000]right:\parbox{5cm}{high wackiness\hfill{}low wackiness}},yshift=.25cm] (thermometer) {\tikz[inner sep=0cm,outer sep=0cm]{
	\def\thermocolors{{0,
		100,200,300,
		400,500,600,
		700,800,900,
	}}
	\foreach[count=\j,evaluate=\c using ({\thermocolors[\j]})] \i in {0,.556,...,5}{
		\def\col{green-}
		\ifthenelse{\j>3}{\def\col{yellow-}}{}
		\ifthenelse{\j>6}{\def\col{red-}}{}
		\node[minimum height=0.556cm,rounded corners=0cm,fill=\col\c] at (0,\i) {};
	}
}};

\draw[-] ([yshift=-.2cm]model2.south) to ([yshift=-.1cm]model2.south) to ([yshift=-.1cm]model1.south west) to (model1.south west);
\draw[-] ([yshift=-.2cm]model2.south) to ([yshift=-.1cm]model2.south) to ([yshift=-.1cm]model3.south east) to (model3.south east);
\node[below=.2cm of model2.south] {Expansion Terms};
\end{tikzpicture}
\caption{Our wackiness score assigns a quantifiable value to each token in a sparse representation, independently of the underlying retrieval model. The three models listed from left to right correspond to one that produces very few wacky tokens to one that produces highly wacky tokens.}
\label{fig:overview}
\end{figure}
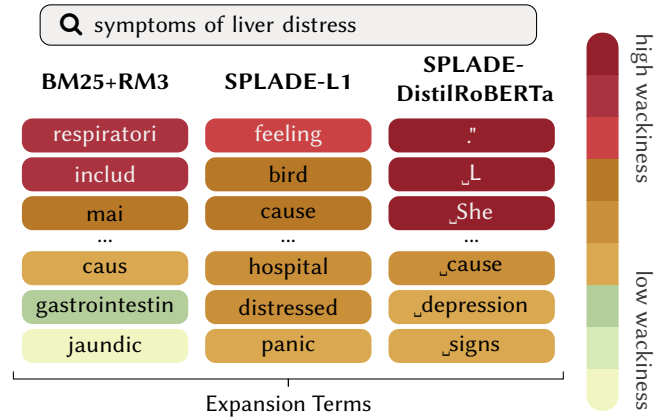

However, the interpretation of these terms is not always straightforward. While many expansion terms are semantically related (e.g. in Figure~\ref{fig:overview}, the query is expanded with terms ``\textvisiblespace{}signs'', ``hospital''), prior qualitative analyses have observed the emergence of wacky weights~\cite{mackenzie2021wacky}: terms that are assigned high importance scores by the model but appear semantically unrelated to the input. Some inputs may be expanded with stopwords, punctuation marks (e.g. ``."'', ``\textvisiblespace{}She'' on Figure~\ref{fig:overview}), or proper names completely unrelated to it (e.g., ``edmund'', or ``roland''). This phenomenon not only raises fundamental questions about the mechanisms of LSR, but calls into question common claims about the intepretability of LSR.

In this paper we perform a reproducibility study of the SPLADE family of models to provide a systematic understanding of the wacky weights phenomenon. We present the first comprehensive dissection of these weights in SPLADE models. We move beyond anecdotal observation to formally define wackiness based on the lexical utility of expansion terms. By quantifying this phenomenon, we trace the origins of these weights and measure their contribution to model effectiveness. Our contributions are as follows:
\begin{enumerate*}[leftmargin=*,label=(\arabic*)]
\item We formally define the wackiness of a token through a lexical \emph{Wackiness Score} to quantitatively compare the wackiness of different tokens. Building on top of this score, we further introduce the \emph{Normalized Wackiness Curve} and the \emph{W-AUC} score to quantitatively compare wackiness across models with different vocabularies and sparsity levels. 
\item We investigate the impact of wacky tokens on retrieval effectiveness for SPLADE-v2 and SPLADE-v3. For both models, wacky tokens are primarily used for in-domain effectiveness rather than out-of-domain generalization. Furthermore, we observe that SPLADE-v3 exhibits a higher dependency on wacky tokens compared to SPLADE-v2.
\item We reproduce SPLADE-v2 across various settings, such as pooling approaches, backbone checkpoints, sparsity regularizers, and fine-tuning datasets. We then systematically evaluate the prevalence of wacky tokens in their outputs, finding that larger vocabulary sizes are associated with a higher prevalence of wacky terms, while stricter sparsity regularizers are associated with a lower prevalence.
\end{enumerate*}
\section{Related Work}
We next provide some background on LSR approaches, interpretability of retrieval models, and the wacky weights phenomenon.

\paragraph{Expansion and Learned Sparse Retrieval}
Traditional retrieval models that rely on lexical matching, such as BM25~\cite{robertson2009probabilistic}, despite their efficiency, suffer from vocabulary mismatch~\cite{furnas1987vocabulary}. Exact-match systems often fail to retrieve relevant documents that do not contain the exact tokens found in the query. To address this, several approaches have proposed expansion mechanisms to bridge the gap between query and document vocabularies. For instance, pseudo-relevance feedback models such as RM3~\cite{jaleel2004umass} use terms from the top initially retrieved documents to expand the query. Additionally, there exist several neural generative expansion approaches, such as Query2Doc~\cite{wang2023query2doc} and Doc2Query~\cite{nogueira2019document, nogueira2019doc2query}.

More recent LSR models unify term weighting and expansion. For instance, DeepCT~\cite{dai2019context}, COIL~\cite{gao2021coil}, uniCOIL~\cite{lin2021few}, and DeepImpact~\cite{mallia2021learning} improve term weighting by using contextualized signals from a transformer encoder~\cite{vaswani2017attention}. However, some of these models either do not perform expansion at all or rely on Doc2Query approaches. SparTerm~\cite{bai2020sparterm} and SPLADE~\cite{formal2021splade} learn term weighting and expansion end-to-end by using the masked language modeling (MLM) head of BERT~\cite{devlin2019bert} to implicitly weight and expand terms into the full vocabulary space, achieving state-of-the-art results without additional models. Since the original SPLADE model, many modifications have been released, e.g., SPLADE-v2 and \mbox{SPLADE-v3}, which introduce architectural improvements and new training schemes that improve both effectiveness and efficiency. Despite these gains, the mechanisms of the learned expansions remains poorly understood. In this work, we investigate the interpretability of LSR, with a focus on expansion terms.

\paragraph{Interpretability of Retrieval Models} 
Interpretability research aims to explain the decision-making process of models in human-underst\-andable terms~\cite{anand2022explainable}. As retrieval models grow in size and complexity, it becomes increasingly important to understand their behavior and to ensure human oversight. 
Modern retrieval architectures such as dense retrievers (e.g. DPR~\cite{karpukhin2020dense}, ANCE~\cite{xiong2020approximate}, TAS-B~\cite{hofstatter2021efficiently}) and cross encoders (e.g. MonoT5~\cite{nogueira2020document}, RankT5~\cite{zhuang2023rankt5}) are generally considered uninterpretable by design. In contrast, LSR models, such as uniCOIL~\cite{lin2021few}, DeepImpact~\cite{mallia2021learning}, EPIC~\cite{macavaney2020expansion}, and SPLADE~\cite{formal2021splade}, project learned weights back into vocabulary space. These models claim interpretability by design, as every dimension in their output corresponds to a natural language token, making retrieval auditable.

In practice, models like SPLADE are not fully interpretable. Prior analysis has documented the phenomenon of wacky weights~\cite{mackenzie2021wacky}, i.e., instances where the model assigns high relevance scores to terms that appear semantically unrelated to the query. In this work, we approach this problem by quantitatively investigating the origins and impact of this phenomenon to better understand why SPLADE's learned expansions diverge from human intuition.

\paragraph{Wacky Weights Phenomenon}
The wacky weights phenomenon was first identified by Mackenzie et al.~\cite{mackenzie2021wacky}. The authors noted that the document-at-a-time (DaaT)~\cite{mallia2017faster, broder2003efficient, ding2011faster} query processing strategy, despite obtaining high effectiveness compared to classical sparse algorithms~\cite{crane2017comparison}, is outperformed by score-at-a-time (SaaT)~\cite{lin2015anytime, trotman2019micro} strategies when applied to learned sparse retrieval models like SPLADE. Moreover, they qualitatively observed that SPLADE-v2 tends to expand queries with non-meaningful stopword tokens assigned high weights. For instance, the query ``androgen receptor define'' was expanded with stopwords such as ``is'' (weight 70), ``the'' (weight 56), and ``for'' (weight 46), and a comma token ``,'' (weight 68). They called these assignments ``wacky weights'' and hypothesized that they are the primary cause of the effectiveness drop of DaaT approaches on LSR models.

To understand whether SPLADE encodes relevance signals within stopword tokens~\citet{mackenzie2023exploring} trained variants of SPLADE-v2 restricted to specific vocabulary subsets. While these restricted versions did not match the effectiveness of the original SPLADE-v2, they still demonstrated strong effectiveness, outperforming the BM25 baseline on the MS MARCO~\cite{nguyen2016ms}. Consequently, the authors concluded that SPLADE has the capacity to assign useful relevance signals to seemingly meaningless tokens. Porco et al.~\cite{porco2025alternative} went one step further, attempting to mitigate the wacky weights problem. They observed that SPLADE models expand many documents using the same small set of stopwords. This behavior results in very long posting lists for these specific tokens, which increases query latency. To address this, the authors proposed the DF-FLOPS loss function, which explicitly constrains the resulting posting list lengths. This modification helped to reduce the number of stopwords in SPLADE representations, drastically reducing the retrieval latency.
 
In short, prior works have either qualitatively observed the wacky weights phenomenon or attempted to mitigate it. Instead, we propose a systematic approach to investigating this phenomenon. We introduce a formal definition of wacky tokens and a numerical score to quantify the degree of wackiness based on this definition. Furthermore, we compare different versions of the SPLADE model in terms of the prevalence of these tokens and analyze their impact on the model's retrieval effectiveness.
\enlargethispage{\baselineskip}
\section{Methodology}
Here we detail how SPLADE is trained and how we modify it, how we identify wacky tokens, and how we measure model wackiness.

\subsection{Model Architecture and Training}
\label{sec:model_architecture}
SPLADE is based on a pre-trained transformer encoder that maps an input sequence of tokens $t = (t_1, \dots, t_L)$ to a sequence of hidden states $h = (h_{t_1}, \dots, h_{t_L})$, i.e., contextualized embeddings. To obtain a sparse representation in the vocabulary space $|V|$, SPLADE applies a projection layer followed by an activation function to each token's final hidden state. Specifically, the final sparse vector $v_{t_i} \in \mathbb{R}^{|V|}$ for the $i$-th token $t_i$ is computed as $v_{t_i} = \log(1 + \text{ReLU}(Wh_{t_i} + b))$,
where $Wh_{t_i} + b$ is the linear unembedding layer output ($W$ is the transpose of the embedding matrix and $b$ is a bias term) and the $\log(1 + \text{ReLU}(\cdot))$ activation ensures non-negative, sparse weights.

To produce the sparse representation for documents ($v_d\in \mathbb{R}^{|V|}$) or queries ($v_q\in \mathbb{R}^{|V|}$), SPLADE aggregates sparse vectors of individual tokens. Most prior works employ maximum aggregation (MAX), which pools the highest weight for each vocabulary dimension across all tokens~\cite{formal2021splade, lassance2024splade, porco2025alternative}. We also experiment with sum aggregation (SUM) and CLS aggregation (CLS) to investigate if the aggregation function could influence the presence of wacky tokens:
\begin{equation*}
v_{q/d}^{(j)}\quad=\quad 
\begin{cases} 
    \quad\displaystyle \max_{t_i \in q/d} v_i^{(j)} & \quad\text{(MAX aggregation)}\\[15pt]
    \quad\displaystyle \sum_{t_i \in q/d} v_i^{(j)} & \quad\text{(SUM aggregation)}\\[15pt]
    \quad{}v_{\text{[CLS]}}^{(j)} & \quad\text{(CLS aggregation)}
\end{cases},
\end{equation*}
where $v_{t_i}^{(j)}$ denotes the weight of the $j$-th term from the vocabulary predicted by the $i$-th token.

To learn these representations, SPLADE optimizes a ranking loss (e.g., InfoNCE~\cite{oord2018representation} or MarginMSE~\cite{hofstatter2020improving}) alongside regularization terms that enforce sparsity. The total loss $\mathcal{L}$ is defined as:
\begin{equation*} 
    \mathcal{L}\enskip=\enskip\mathcal{L}_{\text{rank}}\enskip+\enskip\lambda_q \mathcal{L}_{\text{reg}}^{(q)}\enskip+\enskip\lambda_d \mathcal{L}_{\text{reg}}^{(d)},
\end{equation*}
where $\mathcal{L}_{\text{rank}}$ denotes the ranking loss, $\mathcal{L}_{\text{reg}}^{(q)}$ and $\mathcal{L}_{\text{reg}}^{(d)}$ represent the regularization terms applied to the query and document representations respectively, and $\lambda_q$ and $\lambda_d$ are scalar hyperparameters that control the trade-off between sparsity and retrieval effectiveness. Most prior works use FLOPs loss~\cite{paria2020minimizing} as a regularization term:
\begin{equation*}
\mathcal{L}_{\text{FLOPS}}\enskip=\enskip\sum_{j \in V}~\left(\ \frac{1}{N} \sum_{i=1}^{N} v_{q_i/d_i}^{(j)}\ \right)^2,
\end{equation*}
where $N$ is the batch size and \raisebox{.25\baselineskip}{$v_{q_i/d_i}^{(j)}$} is the weight of a term $j$ in the sparse representation of a document $d_i$ or a query $q_i$. FLOPs penalizes terms that appear frequently across the batch, which results in reducing the lengths of posting lists in the model, improving retrieval efficiency. In this work, we also experiment with L1 regularization to investigate if the loss function influences the presence of wacky tokens. In contrast to FLOPs, L1 operates on individual inputs, forcing the model to make them as sparse as possible:
\begin{equation*}
\mathcal{L}_{\text{L1}}\enskip=\enskip\frac{1}{N} \sum_{i=1}^{N}~\left(\ \sum_{j \in V} |v_{i}^{(j)}|\ \right).
\end{equation*}

\subsection{Identifying Wacky Tokens}\label{sec:identifying_wacky_tokens}
In prior works~\cite{mackenzie2021wacky, mackenzie2023exploring}, wacky weights were primarily identified based on manual examination of model output. Therefore, one of the core challenges of this paper is to formally define ``wackiness'' of a token without relying on subjective human judgment.

We formally define a wacky token based on its lexical importance in the context of retrieved documents. A token is considered wacky if it is
(1) generated by the model during the expansion process (i.e., it is not part of the original input text), and 
(2) has a low lexical importance within the set of documents retrieved by that input. 
Based on this definition, we developed a \emph{Wackiness Score} for each token in the vocabulary. The score is computed according to the following steps.

\paragraph{Expansion}
For a given input text $x$, let $T_x^{\text{orig}} \subset V$ be the set of its original tokens. We compute the sparse representation $v_x$ of $x$ using the model under investigation and identify the set of all activated tokens:
$$
T_x^{\text{model}} = \{t \in V \mid v_x[t] > 0 \}.
$$
The set of expansion tokens is then formally defined as the difference between the model's output tokens and the original lexical tokens:
\begin{equation*}
 T_x^{\text{exp}}\enskip=\enskip{}T_x^{\text{model}} \setminus T_x^{\text{orig}}
\end{equation*}

\paragraph{Retrieval}
We use the sparse representation $v_x$ to retrieve the set of top-k documents, $D_x = \{d_1, \dots, d_k\}$, from the background collection $C$. Note that if the input $x$ is a document, it is treated as a query for this retrieval step to identify the top-k documents. The documents are ranked according to the original ranking function of the model being evaluated.

\paragraph{Lexical Importance Calculation}
For each expansion token \mbox{$t \in T_x^{\text{exp}}$}, we calculate its lexical importance score $S(t, x)$ within the retrieved set $D_x$ using TF-IDF:
\begin{equation*}
S(t, x)\enskip=\enskip\frac{\sum_{d \in D_x} \text{count}(t, d)}{\sum_{d \in D_x} \text{len}(d)}\enskip\cdot\enskip\log \left( \frac{N}{|\{d \in C : t \in d\}|} \right),
\end{equation*}
Here, the TF term is the total frequency of $t$ across all documents in $D_x$, and the IDF term is calculated over the entire background collection $C$ (containing $N$ total documents).

\paragraph{Averaging}
For each token $t \in V$, we compute the average lexical importance across all inputs $X_t$ where $t$ appears as an expansion token. We then normalize these averages to the range $[0, 1]$. The final Wackiness Score is defined as one minus the normalized average:
\begin{equation*}
    \text{WackinessScore}(t)\enskip=\enskip1 - \left( \frac{1}{|X_t|} \sum_{x \in X_t} S(t, x) \right)
\end{equation*}

This definition is universal across any form of expansion method. When analyzing wackiness of query expansion tokens, the input set consists of search queries. When analyzing wackiness of document expansion tokens, the input set consists of documents from the collection. In this case, we treat each document as a query and measure whether the document expansion tokens are lexically important within the set of its nearest neighbors.

\subsection{Comparing Wackiness Across Models}
\label{sec:wackiness_metric}
In order to understand the phenomenon of wacky tokens, we need to not only detect artifacts of expansion but also measure their prevalence to compare models in terms of their wackiness. Thus, any proposed metric must be comparable across different models. This presents a significant challenge due to several factors: first, models differ in sparsity, meaning those that produce more tokens might produce more wacky tokens; second, models use different vocabularies, which makes comparing the wacky scores of similar tokens impossible. To address these challenges, we introduce the \emph{Normalized Wackiness Curve}. The construction of this curve proceeds in three steps.

Firstly, for a given model, we first calculate the \emph{Wackiness Score} for its tokens, as defined in Section~\ref{sec:identifying_wacky_tokens}. We then rank all tokens in descending order based on their score.
Secondly, to normalize for different vocabulary sizes $|V|$, we partition the ranked tokens into $B$ equal-sized bins. The $i$-th bin contains all tokens ranked from index $\left\lfloor \frac{(i) |V|}{B} \right\rfloor$ to $\left\lfloor \frac{(i+1) |V|}{B} \right\rfloor$. This ensures that each bin represents a fixed percentage of the total vocabulary rather than a fixed number of tokens, making scores of models with different vocabulary sizes comparable.
Finally, for each bin, we compute the mean \emph{Wackiness Score} of its tokens.

We then visualize the normalized distribution of wacky tokens by plotting the calculated average score for each bin. A curve that maintains high magnitudes across a significant portion of the bins indicates that the model assigns high wackiness scores to a broader proportion of its vocabulary. This enables us to visually compare different models in terms of their wackiness. Moreover, to compare the models quantitatively, we calculate the \emph{Area Under the Wackiness Curve (W-AUC)}. Models with a higher \emph{W-AUC} tend to use more wacky tokens.
\section{Experimental Setup}
In the following sections, we describe our experimental setup for analyzing the impact wacky tokens have on retrieval effectiveness and how model architecture influences their presence.

\subsection{Impact of Wacky Tokens on Effectiveness}
\label{sec:wacky_tokens_impact_experimental_setup}
First, we investigate the impact of wacky tokens on retrieval effectiveness. To do this, we remove individual wacky tokens from the query expansion and measure the change in retrieval effectiveness using the following steps:

\begin{enumerate}[leftmargin=*]
\item For a given model, we compute Wackiness Scores for its vocabulary tokens using the methodology described in Section~\ref{sec:identifying_wacky_tokens}.
\item We sort tokens in descending order of Wackiness Score (from most wacky to least wacky).
\item We select various thresholds (e.g., top-100, top-1000, top-10000) and remove the corresponding tokens from the set of expansions. We then measure the resulting retrieval effectiveness.
\item We compare the removal of Wacky Tokens against a random baseline. For each threshold, we remove an equivalent number of randomly selected tokens. This process is repeated 10 times to calculate the mean effectiveness drop and standard deviation (std.) under random token removal. We then estimate the 95\% confidence interval for the random baseline as the mean $\pm$ 2~std.
\end{enumerate}

For this experiment, we use the two most common versions of SPLADE: SPLADE-v2~\cite{formal2021splade} and SPLADE-v3~\cite{lassance2024splade}. Given that these models are predominantly fine-tuned on MS MARCO, we use the MS MARCO development set to assess in-domain effectiveness. To test out-of-domain effectiveness, we select ten diverse evaluation sets from the BEIR benchmark~\cite{kamalloo:2024}: ArguAna, Climate-FEVER, DBPedia-Entity, FiQA-2018, NFCorpus, Quora, SCIDOCS, SciFact, TREC-COVID, and Touché-2020. We explicitly exclude large collections, such as NQ and HotpotQA, due to computational constraints given the need to perform an extensive number of retrieval runs. We believe that any behaviors related to large-scale data will be evident in MS MARCO, given its size and representative complexity.

For MS MARCO, we report several evaluation measures: MRR@10, Recall@10, Recall@100, and Recall@1000. For BEIR, we report NDCG@10. We define thresholds for the top wacky tokens using a logarithmic scale, as we are primarily interested in the effects related to the removal of wacky tokens with high wackiness score.

\subsection{Architectural Influences of Wackiness}
\label{sec:wackiness_metric_experimental_setup}
To isolate factors contributing to the prevalence of wacky tokens, we trained several variants of the SPLADE-v2 model from scratch. Unless otherwise noted, models were fine-tuned on the training subset of MS MARCO using MAX aggregation strategy and FLOPs regularization loss (see Section~\ref{sec:model_architecture}). First, we quantitatively evaluate the wackiness of these models using the Normalized Wackiness Curve and the W-AUC metric defined in Section~\ref{sec:wackiness_metric}. Second, we visually analyze the top-ranked wacky tokens to identify qualitative differences across models. Specifically, we analyze the following fine-tuning parameters:

\begin{figure*}[!ht]
\centering
\includegraphics[width=\linewidth]{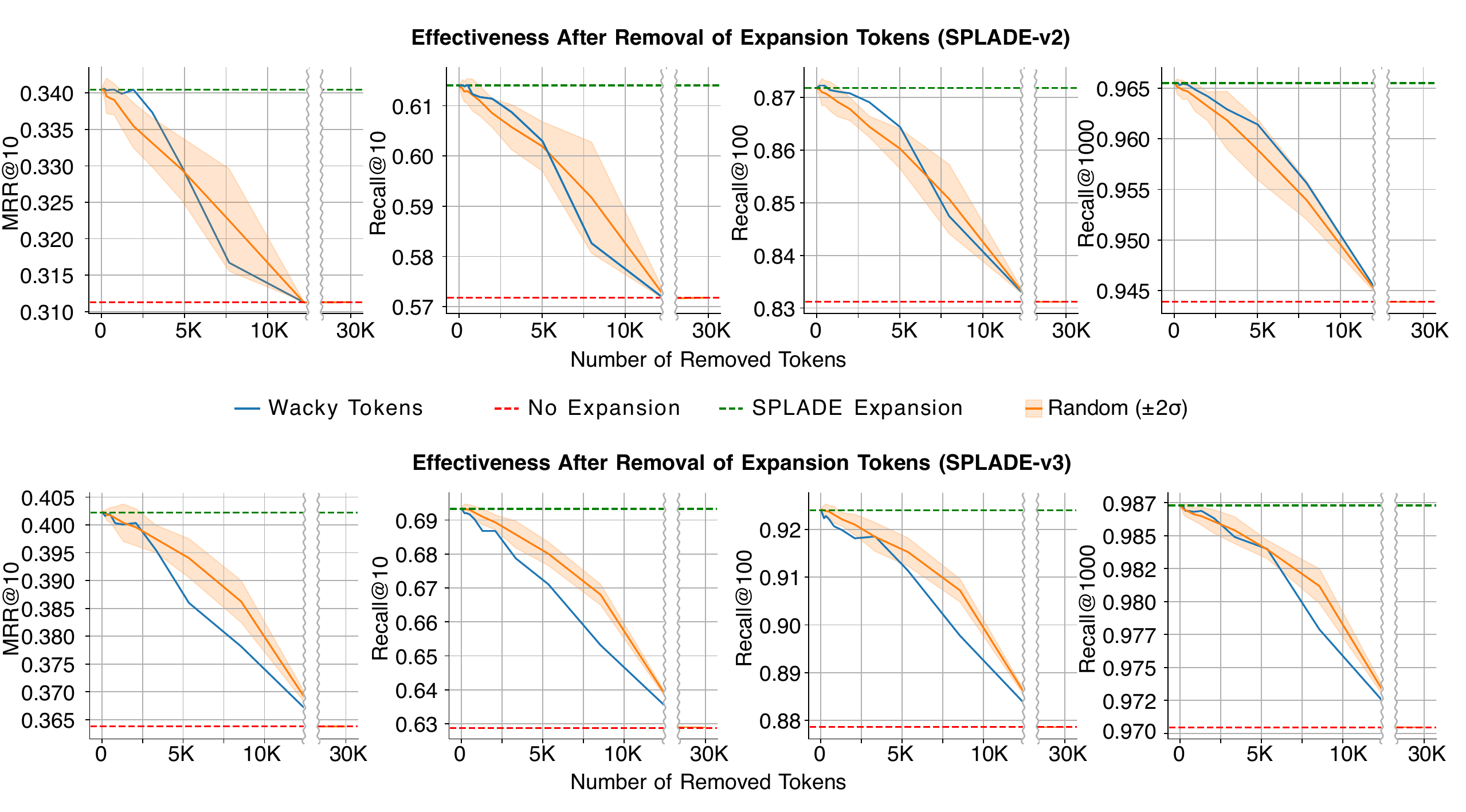}
\caption{Impact of removing top-$N$ wacky tokens vs. random tokens on MS MARCO. The blue line corresponds to effectiveness after removing the top-$N$ wacky tokens. The orange region corresponds to the average $\pm$ two standard deviations of effectiveness after the removal of random expansion tokens. The red and green lines correspond to the cases where no expansion tokens are used for retrieval and where all of the SPLADE expansion tokens are used, respectively.}
\label{fig:msmarco_wacky_tokens_removal}
\end{figure*}
\begin{description}
\item[Base Checkpoint:] We compare the standard DistilBERT initialization against DistilRoBERTa and ModernBERT to investigate if wacky tokens are artifacts of the backbone pre-trained encoder. We denote these models as SPLADE-v2-DistilRoBERTa and SPLADE-v2-ModernBERT.
\item[Training Dataset:] To determine if wacky tokens are artifacts of the fine-tuning dataset, we include a checkpoint fine-tuned on Natural Questions (NQ) instead of MS MARCO. We denote this version as SPLADE-v2-NQ.
\item[Sparsification Loss:] We compare the standard FLOPs regularization loss against an $L_1$ regularization. While FLOPs penalizes frequent terms across the batch to shorten posting lists, $L_1$ operates on individual inputs to enforce representation sparsity. We denote this version as SPLADE-v2-L1.
\item[Aggregation Strategy:] To investigate if wacky tokens are a result of specific choices of aggregation strategy, we train models with SUM and CLS aggregation to compare them against the standard MAX aggregation (see Section~\ref{sec:model_architecture}). We denote these versions as SPLADE-v2-SUM and SPLADE-v2-CLS.
\end{description}
Finally, we compare all SPLADE variants against modular expansion baselines. We define these as systems where input expansion is performed by a standalone component separate from the retrieval model. Using BM25 as the retrieval model, we pair it with three such distinct expansion modules: RM3~\cite{jaleel2004umass}, a classical pseudo-relevance feedback mechanism that expands queries using important terms from top-k retrieved documents; Query2Doc~\cite{wang2023query2doc} which expands queries with generated pseudo-documents using a Large Language Model (LLM); and DocT5Query~\cite{nogueira2019doc2query}, a document expansion approach that appends pseudo-queries to documents prior to indexing using a fine-tuned T5 generative model~\cite{raffel2020exploring}.
For consistency across experiments, we sample a random subset of 10,000 queries and 10,000 documents from the MS MARCO training set. For each model, we perform separate runs on these subsets to obtain the corresponding Wackiness Curves and W-AUC scores.
\section{Results} In the following sections, we analyze the results of reproducing SPLADE to understand the impact of wacky tokens on retrieval effectiveness and how such tokens may arise as a result of model architectural choices. In Section~\ref{sec:effectiveness_impact}, we selectively remove wacky tokens from SPLADE representations to measure their contribution to retrieval effectiveness. In Section~\ref{sec:architectural_influence}, we investigate the origins of this phenomenon, analyzing how various model configurations influence the prevalence and nature of wacky tokens.

\subsection{Impact of Wacky Tokens on Effectiveness}
\label{sec:effectiveness_impact}
We analyze the impact of wacky tokens on retrieval effectiveness from two perspectives: in-domain (i.e., models trained and tested with the same document collection) and out-of-domain (i.e., models trained with one document collection and tested on another).

\paragraph{In-domain Results}
\begin{figure*}[!t]
\centering
\includegraphics[width=\linewidth]{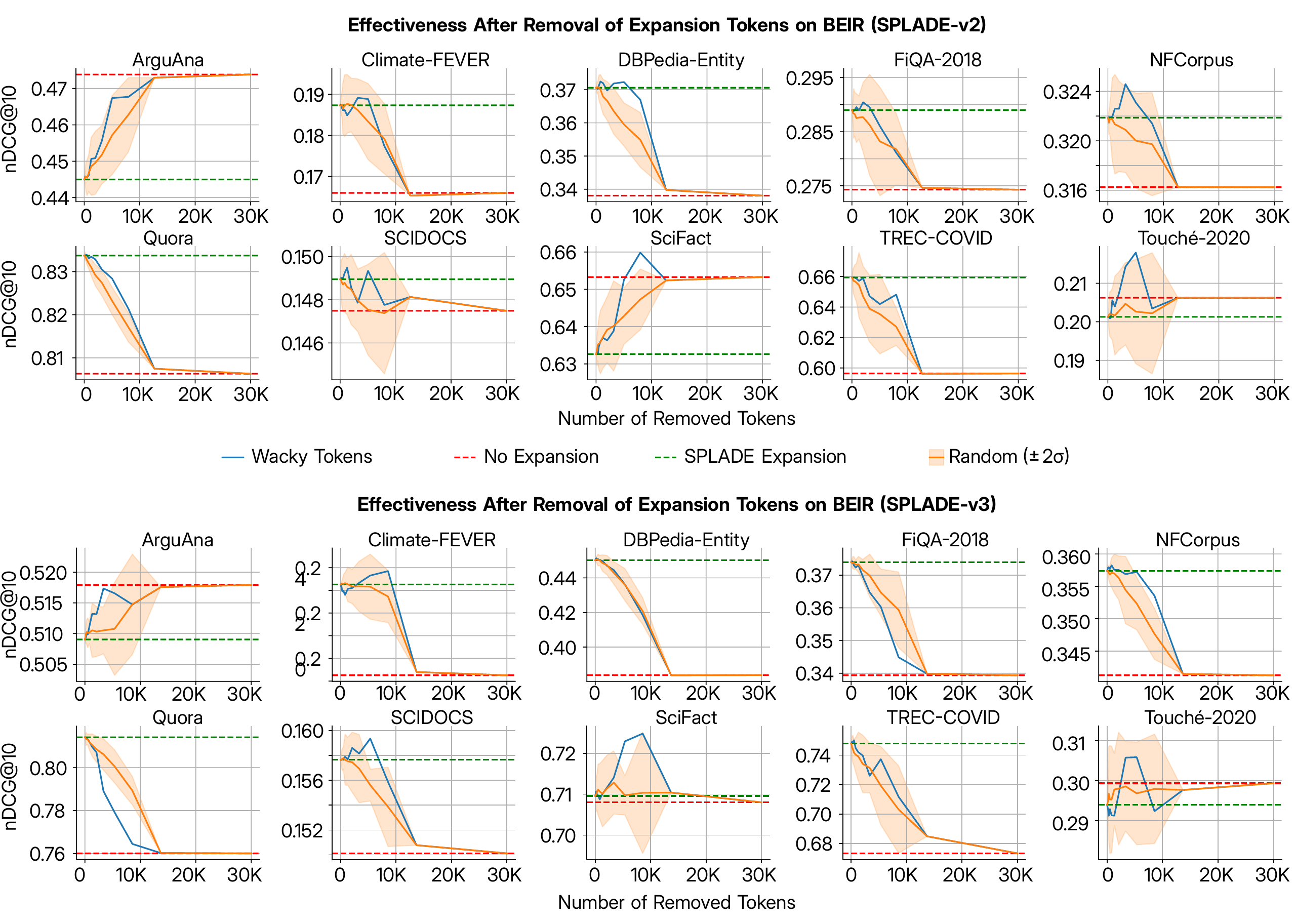}
\caption{Impact of removing top-$N$ wacky tokens vs.\ random tokens on BEIR. The blue lines correspond to effectiveness after removing the top-$N$ wacky tokens. The orange regions correspond to the average $\pm$ two standard deviations of effectiveness after removing random expansion tokens. The red and green lines correspond to the cases where no expansion tokens are used for retrieval and where all of the SPLADE expansion tokens are used, respectively.}
\label{fig:beir_wacky_tokens_removal}
\end{figure*}

Figure~\ref{fig:msmarco_wacky_tokens_removal} illustrates the retrieval effectiveness of SPLADE-v3 and SPLADE-v2 on the MS MARCO development set after the removal of the top-ranked wacky tokens compared to randomly removing expansion tokens. 

Overall, removing a small number of wacky tokens (up to \num{1000}) is either indistinguishable from removing random tokens (in the case of SPLADE-v3) or results in slightly better retrieval effectiveness than the random baseline (in the case of SPLADE-v2). 
However, as more tokens are removed, the two models start to behave differently. For SPLADE-v3, removing many tokens with a high Wackiness Score results in a considerable degradation of retrieval effectiveness that falls outside of the random baseline 95\% confidence interval. In contrast, for SPLADE-v2, removing many tokens with high Wackiness Score remains statistically indistinguishable from removing random tokens. First, these results suggest that SPLADE-v3 exhibits a substantially stronger reliance on wacky tokens compared to SPLADE-v2. Second, the substantial drop in SPLADE-v3 confirms that these tokens are not useless artifacts, but rather an integral part of the model's internal logic, used to encode concepts that extend beyond semantic matching.
Furthermore, we observe that for SPLADE-v3, the effectiveness drops associated with removing wacky tokens are most pronounced in Recall@10, whereas for Recall@1000, they are almost indistinguishable from the removal of random expansion tokens. We hypothesize that this is because the model uses wacky tokens to make fine-grained adjustments to the ranking of top documents, rather than relying on them for the initial retrieval stage.
Overall, these findings highlight that the reliance on wacky tokens can vary across different model architectures. In subsequent sections, we investigate which specific parameters influence it.

\paragraph{Out-of-domain Results} 
Figure~\ref{fig:beir_wacky_tokens_removal} illustrates the retrieval effectiveness of  SPLADE-v2 and SPLADE-v3 on various BEIR datasets after the removal of the top-ranked wacky tokens compared to randomly removing expansion tokens.
Overall, on most datasets, the effectiveness drops associated with removing wacky tokens are either indistinguishable or slightly smaller than those associated with removing random expansion tokens. The only exception is on the Quora test collection, where the models exhibit patterns similar to their in-domain effectiveness. 

These results suggest that wacky tokens play a more significant role for in-domain retrieval effectiveness, which indicates that models use these tokens to specifically adapt to the dataset they were fine-tuned on. 
Combined with the previous observations, specifically the fact that models use wacky tokens for fine-grained adjustments of rankings, we hypothesize that wacky weights function primarily as a mechanism for dataset-specific overfitting. Together, these results suggest that models use these tokens to store specialized ranking signals that extend beyond semantic matching.

\subsection{Architectural Influences of Wackiness}
\label{sec:architectural_influence}
\begin{table}[t!]
\sffamily
\centering
\setlength{\tabcolsep}{0pt}
\renewcommand{\arraystretch}{1.2}
\caption{Retrieval effectiveness of expansion baselines, trained SPLADE variants evaluated on the MS MARCO development set, TREC DL 19, and TREC DL 20. $^*$ denotes statistical significance with $p < 0.05$ and $^{**}$ with $p < 0.01$ compared to reproduced SPLADE-v2 (Repro.) using a paired t-test with Bonferroni correction.}\label{tab:splade_variants_performance}
\newcommand{\rcw}{3em}
\begin{tabular*}{\linewidth}{@{\extracolsep{\fill}}lp{\rcw}p{\rcw}@{}p{\rcw}p{\rcw}@{}p{\rcw}p{\rcw}}
\toprule
 & \multicolumn{2}{c}{\textbf{MS MARCO}} & \multicolumn{2}{c}{\textbf{TREC DL 19}} & \multicolumn{2}{c}{\textbf{TREC DL 20}} \\
\cmidrule(lr){2-3} \cmidrule(lr){4-5} \cmidrule(lr){6-7}
\textbf{Model Variant} & \centering\rotatebox{90}{MRR@10} & \centering\rotatebox{90}{R@1000} & \centering\rotatebox{90}{nDCG@10} & \centering\rotatebox{90}{R@1000} & \centering\rotatebox{90}{nDCG@10} & \centering\rotatebox{90}{R@1000}\arraybackslash \\
\midrule
\multicolumn{5}{l}{\textit{Expansion Baselines}} \\
BM25+RM3 & 0.194$^{**}$ & 0.877$^{**}$ & 0.526$^*$ & 0.767 & 0.509$^{**}$ & 0.776 \\
BM25+Query2Doc & 0.231$^{**}$ & 0.923$^{**}$ & 0.659 & 0.842 & 0.650 & 0.823 \\
BM25+DocT5Query & 0.275$^{**}$ & 0.921$^{**}$ & 0.587 & 0.686 & 0.599 & 0.715 \\[.5em]
\midrule
\multicolumn{5}{l}{\textit{SPLADE Baselines}} \\
SPLADE-v3 & 0.403$^{**}$ & 0.987$^{**}$ & 0.722 & 0.879$^{**}$ & 0.751$^{**}$ & 0.844$^{**}$ \\
SPLADE-v2 & 0.340 & 0.966 & 0.683 & 0.826 & 0.672 & 0.755 \\
SPLADE-v2 (Repro.) & 0.342 & 0.966 & 0.671 & 0.810 & 0.665 & 0.754 \\[.5em]
\midrule
\multicolumn{5}{l}{\textit{Different Sparsification Loss}} \\
SPLADE-v2-L1 & 0.338 & 0.963 & 0.698 & 0.806 & 0.653 & 0.730 \\[.5em]
\midrule
\multicolumn{5}{l}{\textit{Different Base Checkpoint}} \\
v2-DistilRoBERTa & 0.302$^{**}$ & 0.960 & 0.647 & 0.793 & 0.614 & 0.685$^*$ \\
v2-ModernBERT & 0.340 & 0.974$^{**}$ & 0.655 & 0.792 & 0.660 & 0.712 \\[.5em]
\midrule
\multicolumn{5}{l}{\textit{Different Dataset}} \\
SPLADE-v2-NQ & 0.187$^{**}$ & 0.871$^{**}$ & 0.475$^{**}$ & 0.692$^{**}$ & 0.399$^{**}$ & 0.626$^{**}$ \\[.5em]
\midrule
\multicolumn{5}{l}{\textit{Different Aggregation Function}} \\
SPLADE-v2-SUM & 0.316$^{**}$ & 0.952$^{**}$ & 0.664 & 0.797 & 0.619 & 0.723 \\
SPLADE-v2-CLS & 0.269$^{**}$ & 0.893$^{**}$ & 0.560$^*$ & 0.635$^{**}$ & 0.560$^{**}$ & 0.556$^{**}$ \\
\bottomrule
\end{tabular*}
\end{table}
In the previous section, we observed that the influence of wacky tokens on retrieval effectiveness varies between the two tested models, SPLADE-v2 and SPLADE-v3. In this section, we investigate whether specific training strategies, such as the choice of the base pre-trained encoder, the fine-tuning dataset, and the loss function, influence the prevalence of these tokens in the model output.

\paragraph{Retrieval Effectiveness}
\begin{figure}[t]
\includegraphics[width=\linewidth]{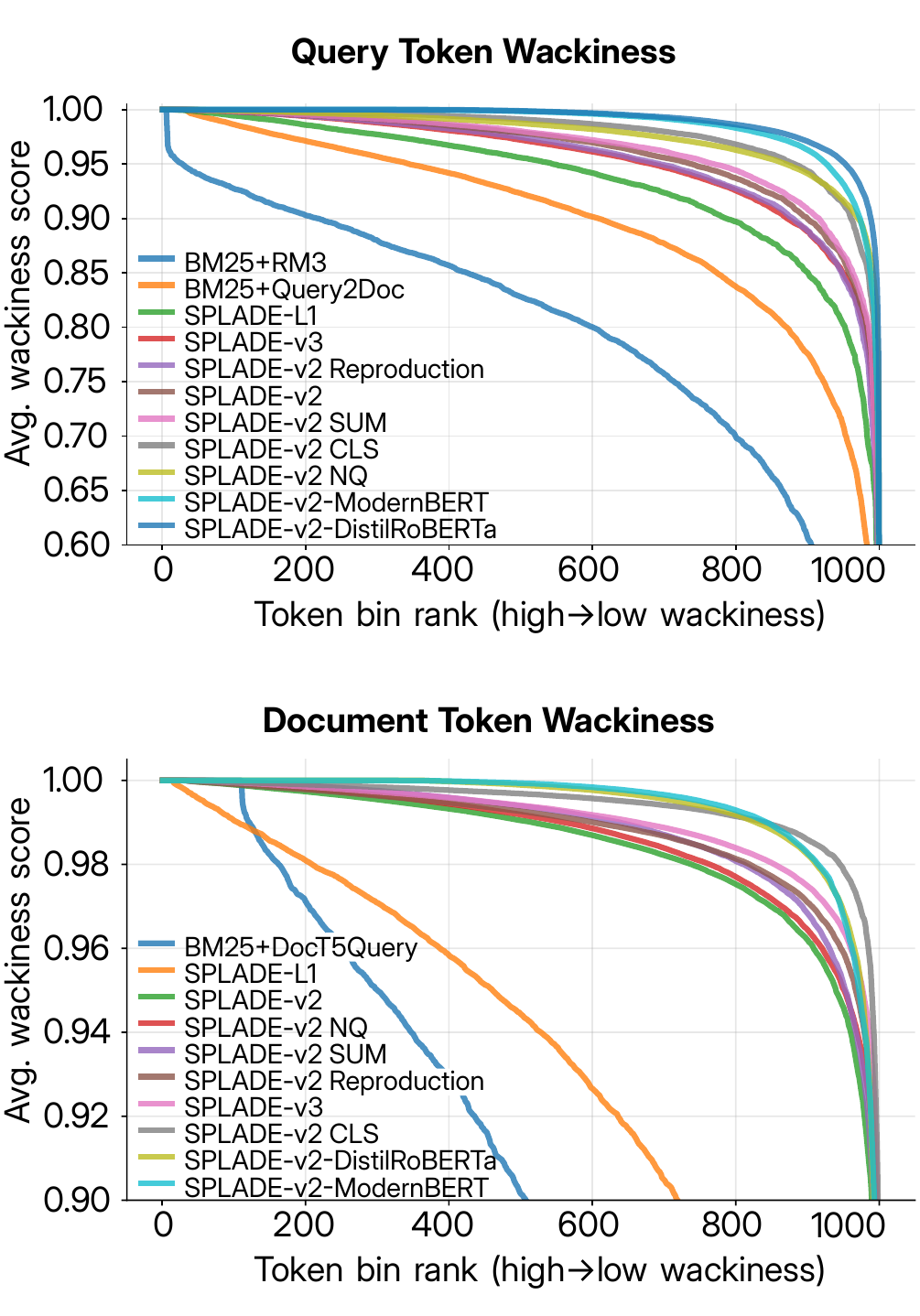}
\caption{Normalized Wackiness Curves for different versions of SPLADE as well as expansion baselines (BM25 + RM3, BM25 + Query2Doc, BM25 + DocT5Query). \textit{Top}: Wackiness curves calculated for query expansion tokens. \textit{Bottom}: Wackiness curves calculated for document expansion tokens. A shallower curve corresponds to a lower W-AUC score, indicating a model with a lower prevalence of wacky tokens.}
\label{fig:token_wackieness_comparison_across_models}
\end{figure}

We start by analyzing the retrieval effectiveness of all considered models. Table~\ref{tab:splade_variants_performance} presents the retrieval effectiveness results on the MS MARCO development set, and the TREC DL 19 and TREC DL 20 evaluation sets. For MS MARCO, we report MRR@10 and Recall@1000. For both TREC DL datasets, we report NDCG@10 and Recall@1000. First, all SPLADE baselines are considerably more effective than the BM25-based expansion baselines (BM25+RM3, BM25+Query2Doc, and BM25+DocT5Query). These results indicate that end-to-end trained SPLADE models are more effective at capturing retrieval signals than modular expansion methods. Second, we successfully replicated the retrieval effectiveness of the original SPLADE-v2, closely matching measures originally reported on MS MARCO and obtaining similar results on TREC DL. These findings are confirmed by a two-tailed paired t-test, which showed no statistically significant difference between the results of the reproduced model and the original one ($p > 0.05$). Third, most SPLADE variants using different backbone models, aggregation functions, or training datasets obtain lower effectiveness compared to the standard SPLADE configurations. These results suggest that the default configurations proposed by the original authors remain highly effective. The only variants with retrieval effectiveness approaching the baselines are SPLADE-v2-ModernBERT, which uses the ModernBERT backbone transformer, and SPLADE-v2-L1, trained with $L_1$ sparsification loss. We attribute the lower effectiveness of the remaining variants to the following key factors. First, SPLADE-v2-NQ likely suffers from the domain shift problem. Trained solely on Natural Questions dataset, it is difficult for the model to generalize to MS MARCO queries and passages. Second, the lower effectiveness of SPLADE-v2-CLS and SPLADE-v2-SUM is likely inherent to the nature of the aggregation functions. SPLADE-v2-CLS forces the model to condense all input information into the [CLS] token, thereby losing the fine-grained representational power of individual tokens. Meanwhile, the lower effectiveness of SPLADE-v2-SUM aligns with the findings of the original SPLADE-v2 work~\cite{formal2021splade}. The SUM aggregation function depends highly on the length of the input and could propagate noise to the final output. In the following sections we further discuss these results in terms of the prevalence of wacky tokens in expansion outputs of these models.

\paragraph{Quantitative Results}

Figure~\ref{fig:token_wackieness_comparison_across_models} illustrates the Normalized Wackiness Curves for all considered models (described in Sections \ref{sec:wackiness_metric} and \ref{sec:wackiness_metric_experimental_setup}). Figure~\ref{fig:token_wackieness_comparison_across_models} (above) represents Normalized Wackiness Curves calculated based on query expansions, while Figure~\ref{fig:token_wackieness_comparison_across_models} (below) represents the Normalized Wackiness Curves calculated based on document expansions. While the query-based Wackiness Curves show distinct model behaviors, the document-based curves are less informative. We believe this to be due to a lack of homogeneity in the topical distribution of MS MARCO passages, meaning there are likely only few similar related passages to the target passages. In the case of the query curve, the relevant document is often a source of lexical importance. This topical distribution problem with MS MARCO artificially inflates the wackiness of all models, making them hard to distinguish, in contrast to the observations on the query expansion plot.

Table~\ref{tab:model_wackiness_auc_scores} presents the corresponding Area Under the Wackiness Curve (W-AUC) scores. A shallower curve corresponds to a lower W-AUC score, indicating a model with a lower prevalence of wacky tokens. Overall, the statistical expansion baseline (BM25+RM3) exhibits the lowest prevalence of wacky tokens among all models. This finding is consistent with prior work~\cite{mackenzie2021wacky} and aligns with the intuition that RM3 expands queries using terms explicitly found in the top initially retrieved documents, thus naturally minimizing the divergence between expansion terms and the document vocabulary.

\begin{table}[t!]
\sffamily
\centering
\caption{Sorted W-AUC scores for all considered versions of SPLADE and BM25 + RM3. Higher W-AUC indicates higher prevalence of wacky tokens in the model output. Superscript letters denote statistical significance groupings (models sharing a letter are not significantly different via paired t-test with Bonferroni correction over 10 runs, $p \ge 0.05$).}
\renewcommand{\arraystretch}{1.2}
\begin{tabular*}{\linewidth}{@{\extracolsep{\fill}}ll}
\toprule
\textbf{Model} & \textbf{AUC ± $2\sigma$} $(\downarrow)$ \\
\midrule
BM25+RM3 & 0.833 ± 0.057$^{\text{a}}$ \\
BM25+Query2Doc & 0.890 ± 0.023$^{\text{a}}$ \\
SPLADE-v2-L1 & 0.934 ± 0.002$^{\text{b}}$ \\
SPLADE-v3 & 0.947 ± 0.033$^{\text{bcdg}}$ \\
SPLADE-v2 (Repro.) & 0.953 ± 0.001$^{\text{c}}$ \\
SPLADE-v2 & 0.959 ± 0.002$^{\text{d}}$ \\
SPLADE-v2-SUM & 0.962 ± 0.001$^{\text{e}}$ \\
SPLADE-v2-NQ & 0.976 ± <0.001$^{\text{f}}$ \\
SPLADE-v2-CLS & 0.977 ± 0.001$^{\text{efg}}$ \\
SPLADE-v2-ModernBERT & 0.985 ± 0.001$^{\text{h}}$ \\
SPLADE-v2-DistilRoBERTa & 0.989 ± <0.001$^{\text{j}}$ \\
\bottomrule
\end{tabular*}
\label{tab:model_wackiness_auc_scores}
\end{table}

Furthermore, all modular expansion baselines (BM25+Query2Doc and BM25+DocT5Query), which rely on neural architectures, are more wacky than the statistical BM25+RM3 but less wacky than all considered SPLADE versions. This is expected, as these models are trained to provide natural, semantically related expansions.

Among the SPLADE variants, several key trends emerge. First, both Figure~\ref{fig:token_wackieness_comparison_across_models} and Table~\ref{tab:model_wackiness_auc_scores} indicates that SPLADE-v2-L1 exhibits the lowest prevalence of wacky tokens. This is likely related to the differences between the FLOPs and L1 losses described in Section~\ref{sec:model_architecture}. The FLOPs loss is designed to ensure that no terms accumulate excessively long posting lists in the final index, which may still allow the model to store some relevance signals in apparently meaningless tokens. In contrast, the L1 loss imposes a strict sparsity constraint on individual input representations, forcing the model to use more semantically meaningful terms to encode the relevance signal.

\begin{table}[t!]
\sffamily
\centering
\caption{Top-20 wacky tokens across different SPLADE variants as reproduced in this paper. Tokens are sorted by their Wackiness Score.}
\label{tab:wacky_tokens_qualitative_comparison}
\renewcommand{\arraystretch}{1.2}
\setlength{\tabcolsep}{4pt}
\begin{tabular*}{\linewidth}{@{\extracolsep{\fill}}llll}
\toprule
\textbf{\begin{tabular}[l]{@{}l@{}}SPLADE-\\ v2\end{tabular}} & 
\textbf{\begin{tabular}[l]{@{}l@{}}SPLADE-\\ v2-L1\end{tabular}} & 
\textbf{\begin{tabular}[l]{@{}l@{}}SPLADE-\\ v2-NQ\end{tabular}} & 
\textbf{\begin{tabular}[l]{@{}l@{}}SPLADE-v2-\\ ModernBERT\end{tabular}} \\
\midrule
\#\#{\fontencoding{T2A}\selectfont б} & [CLS] & \#\#{\fontencoding{T2A}\selectfont б} & D \\
{}[CLS] & mira & \#\#$^+$ & M \\
... & $\beta$ & \#\#{\fontencoding{T2A}\selectfont д} &  \textvisiblespace{} M \\
• & $\varphi$ & \#\#\cjRL{d} & \textvisiblespace{} D \\
edmund & fischer & " & \textvisiblespace{} It \\
$-$ & hagen & pinned & \textvisiblespace{} This \\
janet & sophia & $^\circ$c & \textvisiblespace{} For \\
mira & gunner & \#\#{\fontencoding{T2A}\selectfont о} & \textvisiblespace{} And \\
santo & roland & \cjRL{k} & AD \\
orthodox & kellan & \#\#o & DE \\
$\varphi$ & rye & \#\#{\fontencoding{T2A}\selectfont п} & So \\
celia & yuri & co$_2$ & \textvisiblespace{} More \\
caleb & ellie & \RL{q} & \textvisiblespace{} Med \\
winkler & 1985 & \#\#\RL{q} & \textvisiblespace{} Now \\
\#\#urian & whilst & \#\#o$\varsigma$ & Now \\
$'$ & tamara & {\fontencoding{T2A}\selectfont з} & Me \\
hagen & catcher & \#\#{\fontencoding{T2A}\selectfont у} & \textvisiblespace{} Table \\
{\fontencoding{T2A}\selectfont в} & edmund & $\pi$ & Ste \\
rafe & eureka & $^1$ & DM \\
,, & fcc & \RL{r} & HD \\
\bottomrule
\end{tabular*}%
\end{table}

Second, we observe that models with larger vocabularies (SPLADE-v2-ModernBERT and SPLADE-v2-DistilRoBERTa), which have approximately \num{50 000} tokens compared to \num{30 000} tokens in the original SPLADE DistilBERT backbone, exhibit the highest prevalence of wacky tokens. This suggests that a larger vocabulary space provides the model with more degrees of freedom, allowing it to distribute relevance signals to semantically unrelated tokens.

Additionally, in both cases, the SPLADE-v2-CLS model exhibits one of the highest wackiness scores. We hypothesize that models store a general sense of meaning about the input inside wacky tokens. As we describe in our qualitative results below, we performed an ad hoc analysis of wacky token meanings and observed that tokens relating to proper name like ``Pedro'' could encode concepts related to Latin America. This relationship was also observed for other name--concept pairs. Thus, it is expected that a model trained solely using the CLS token vector would rely on these terms more heavily, as models typically aggregate general sequence information into the CLS token representation. In contrast, using the aggregated representations of all tokens (as in SPLADE-v2, SPLADE-v3, and SPLADE-v2-SUM) encourages the model to select expansion terms that act as direct synonyms for individual input words

Finally, in both cases SPLADE-v2, SPLADE-v2-SUM, and SPLADE-v3 show similar prevalence of wacky tokens, which is notable for SPLADE-v3. Despite using a comparable amount of wacky tokens, SPLADE-v3 suffers considerable reductions in effectiveness when they are removed, whereas SPLADE-v2 does not.

\paragraph{Qualitative Results}
Finally, we qualitatively analyze the wacky tokens in model outputs. Table~\ref{tab:wacky_tokens_qualitative_comparison} presents the top-20 tokens with the highest Wackiness Scores across four distinct SPLADE variants: SPLADE-v2, SPLADE-v2-L1, SPLADE-v2-NQ, and SPLADE-v2-ModernBERT.

We observe that models with minor architectural differences (SPLADE-v2 and SPLADE-v2-L1) exhibit a consistent pattern, where both models primarily assign high wackiness scores to proper names (e.g., \textit{edmund}, \textit{mira}, \textit{fischer}). However, more substantial architectural modifications lead to drastic shifts in terms of which tokens are used to represent queries and documents. For instance, changing the fine-tuning dataset to Natural Questions (SPLADE-v2-NQ) results in non-Latin tokens, such as Cyrillic, Hebrew, and Arabic characters. Similarly, altering the backbone to ModernBERT (SPLADE-v2-ModernBERT) produces a set of wacky tokens characterized by tokenizer-specific artifacts, such as single capital letters and tokens prefixed with whitespace markers.

These observations indicate that substantial architectural differences and training configurations substantially alter the nature of the wacky tokens. Moreover, the fact that the model trained on Natural Questions exhibits a completely different set of top wacky tokens compared to the MS MARCO models, despite sharing the exact same architecture, strongly suggests that models exploit wacky tokens to adapt to the specifics of the fine-tuning dataset. We leave an investigation into how training data affects wacky tokens for future work.

\section{Conclusion}
In this paper, we conducted a reproducibility study of the SPLADE family of models, with a specific focus on the wacky weights phenomenon---the appearance of expansion terms that lack semantic alignment with the input text. While prior research only anecdotally observed the presence of this phenomenon, our work provides a formal framework to quantify and analyze this behavior. In this work we introduced two numerical scores: \emph{Wackiness Score} to quantify the wackiness of individual tokens within the vocabulary for each model, and the \emph{Normalized Wackiness Curve} along with the \emph{W-AUC} score to compare models in terms of prevalence of wacky tokens in their output. These scores allowed us to move beyond subjective and anecdotal observation and to provide a comparison of this phenomenon across models with different architectures, vocabularies, and sparsity levels. 

Our systematic investigation into the impact and prevalence of wacky tokens by reproducing the data and architectural components of SPLADE training setups yielded three key findings:

\begin{description}
\item[Finding 1:] Wacky tokens are not a useless artifact, and SPLADE models use them primarily for in-domain effectiveness rather than out-of-domain generalization. We observed that removing wacky tokens resulted in a substantial drop in retrieval effectiveness on the training domain (MS MARCO), particularly for SPLADE-v3. However, on out-of-domain evaluation sets (BEIR), the removal of wacky tokens had negligible impact, often indistinguishable from removing random tokens. 
\item[Finding 2:] The sparsification loss and the vocabulary size are the primarily factors influencing the presence of wacky tokens in the SPLADE output. Specifically, we found that models initialized with larger vocabularies (such as ModernBERT or DistilRoBERTa) exhibit a higher prevalence of wacky tokens, as the larger token space provides the model with more degrees of freedom to distribute relevance signals. Conversely, employing $L_1$ regularization instead of the standard FLOPs loss imposes a stricter sparsity constraint on individual inputs, which forces the model to use more semantically meaningful terms and significantly reduces the prevalence of wacky weights.
\item[Finding 3:] Qualitatively, while models with slight architectural differences (SPLADE-v2, SPLADE-v2-L1, SPLADE-v2-SUM) tend to expand inputs with proper names, stopword tokens and punctuation marks, more drastical architectural changes, such as the fine-tuning dataset (SPLADE-v2-NQ) or backbone transformer (SPLADE-v2-ModernBERT), result in a shift towards different categories of tokens such as non-Latin tokens or tokenizer artifacts.
\end{description}

Overall, our work challenges the common assumption that learned sparse retrieval models are interpretable by design. Wacky tokens, while being an integral part of SPLADE's state-of-the-art retrieval effectiveness, lack connection with human-aligned semantics. We believe that this work provides a foundation for future research to develop LSR models that can balance high retrieval effectiveness with genuine interpretability. 

Future work could design novel training objectives that force LSR models to output interpretable terms. Moreover, we plan to investigate the latent topical information encoded in wacky weights, aiming to uncover the functional roles of specific expansion tokens. Finally, the relationship between wacky terms and retrieval effectiveness warrants further investigation. For instance, future research could develop novel pruning techniques that selectively remove low-utility expansions, optimizing the trade-off between retrieval effectiveness and efficiency.

\begin{acks}
The authors thank the International Max Planck Research School for Intelligent Systems (IMPRS-IS) for their support.
\end{acks}
\balance
\bibliographystyle{ACM-Reference-Format}
\bibliography{sigir26-uww-lit}
\end{document}